\documentstyle[12pt, epsf]{article}
\newcommand{\be}{\beta}

\newcommand{\ep}{\epsilon}

\def\be{\begin{equation}}
\def\ee{\end{equation}}
\begin{document}
\begin{center}
\LARGE\bf{On searches for gravitational waves from mini creation event by laser interferometric detectors} \\ 
\vspace{10mm}
\normalsize{Bhim Prasad Sarmah$^{1,2}$, \, S.K. Banerjee$^{1,3}$, \, S.V. Dhurandhar$^1$ and J.V. Narlikar$^1$} \\ 
\small{$1$. Inter-University Centre for Astronomy and Astrophysics,
Post Bag 4, Ganeshkhind \\ Pune-411007, India} \\ 
\small{$2$. Department of Mathematical Sciences, Tezpur University \\ Tezpur-784028, India} \\
\small{$3$. Department of Mathematics, Amity School of Engineering, Sector-125 \\ Noida-201301, India} \\
E-mail : bhim@tezu.ernet.in,~~skb@iucaa.ernet.in,~~sanjeev@iucaa.ernet.in,~~jvn@iucaa.ernet.in \\
\vspace{10mm}
\normalsize\bf{Abstract} \\ 
\end{center}
As an alternative view to the standard big bang cosmology the quasi-steady state cosmology(QSSC) argues that the universe 
was not created in a single great explosion; it neither had a beginning nor will it ever come to an end. The creation 
of new matter in the universe is a regular feature occurring through finite explosive events. Each creation event is called 
a mini-bang or, a mini creation event(MCE). Gravitational waves are expected to be generated due to any anisotropy present 
in this process of creation. Mini creation event ejecting matter in two oppositely directed jets is thus a source of 
gravitational waves which can in principle be detected by laser interferometric detectors. In the present work we consider the gravitational 
waveforms propagated by linear jets and then estimate the response of laser interferometric detectors like LIGO and LISA. 

\section{Introduction}
The Quasi-Steady State Cosmology(QSSC) was proposed and explored in a series of papers by Hoyle et al(1993, 1994a,b, 1995) 
as a possible alternative to the standard big bang cosmology. The QSSC is based on a Machian theory of gravity which 
satisfies the Weyl postulate and the cosmological principle. The effective field equations are Einstein's equations of 
general relativity together with a negative cosmological constant and a trace-free zero mass scalar field which yield a 
wide range of solutions for the spatial sections of zero, positive and negative curvatures(Sachs et al 1996). \\  

Instead of a single initial infinite explosion called the big bang, the QSSC has a universe without a beginning and its dynamical 
behaviour is sustained by an endless chain of mini-bangs better known as mini creation events(MCE) randomly distributed 
over space. The universe itself has a longterm de-Sitter type expansion with a characteristic time scale 
of $\sim 10^{12}$ yrs, along with short term oscillations of period $\sim 50$ Gyrs. The oscillations respond in phase to 
`on-off' creation activity in MCEs, with  matter being created only in strong gravitational fields associated 
with dense aggregates of matter. The typical mini creation event may explain the outpouring of matter and radiation from a 
wide range of extragalactic objects of varying sizes ranging from superclusters-size mass $\sim 10^{16} M_{\odot}$ to masses 
of the order of $10^6$-$10^{13} M_{\odot}$(Hoyle et al 1993). The cosmology has offered alternative interpretations of phenomena 
like the microwave background, abundances of light neclei, the $m-z$ relation of high redshift supernovae, etc. 
(Hoyle, et al. 2000, Narlikar et al. 2002) and has also suggested tests to distinguish it from the standard model 
(Narlikar and Padmanabhan, 2001). One such possibility is provided by gravitational wave astronomy. \\ 

Gravational waves are expected to be generated if anisotropy is present in a mini creation event. Das Gupta and Narlikar(1993) had performed a preliminary calculation 
relating the size and anisotropy of a typical MCE to the feasibility of its being detected by LIGO type detectors. Here we 
carry out a more refined study of MCEs in which matter is ejected more like a jet. We must have a realistic model of the 
detector noise $n(t)$ to decide what information could be extracted from gravitational waveforms. This noise might have both 
Gaussian and non-Gaussian components but we will restrict ourselves to the statistical errors arising from Gaussian noise only. 
We can describe the remaining Gaussian noise by its spectral density $S_n(f)$, where $f$ is the frequency. The form of $S_n(f)$ 
of course depends on the parameters of the detector. \\  

We first consider the cosmogony of the creation process which leads to the creation and ejection of matter. We show why 
the creation phenomenon may have a non-isotropic character, with ejection taking place along preferred directions. To 
measure its gravitational wave effect we perform the calculation of gravitational amplitude generated by mini creation 
event of mass $M$ at a cosmological distance $r$ and then compare the signal to noise ratio [SNR] detectable by the ground based laser 
interferometric detector of the LIGO and the advance LIGO type within the frequency range of $10 Hz - 1000 Hz$ 
with the SNR detectable by the LISA-Laser Interferometric Space Antenna within the frequency range from $10^{-4} Hz$ to $10^{-1} Hz$.

\section{Matter Creation in the QSSC}
\bigskip

The creation of matter in the QSSC proceeds via an exchange of energy from a background reservoir of a scalar 
field $C$ of negative energy and stresses.  The details of the process have been described in Hoyle et al (2000) and 
QSSC papers eg. Hoyle, et al. (1995), Sachs et al. (1996).  We outline here the relevant aspects of the process that concerns us here. \\ 

First, the basic particle to be created is the so called Planck particle which has mass

\begin{equation}
m_p = \sqrt{\frac{3\hbar c}{4\pi G}}
\end{equation}

\noindent This particle itself is unstable on the time scale of $10^{-43}$ seconds. It subsequently decays into baryons through 
a series of processes that are currently discussed in the GUTs $\rightarrow$ Quark Gluon Plasma $\rightarrow$ Baryons in 
high energy particle physics. However, the initial stage of this sequence of events concerns us here, viz. the location of 
creation of Planck particles.  Can creation take place anywhere?
The answer is `No'.  The process requires a high enough energy threshold of the $C$ - field:
\begin{equation}
C_iC^i = m_p^2c^4,
\end{equation}

\noindent where $C_i = \partial C/\partial x^i $, $x^i$ ($i = 0,1,2,3, x^0$ time like) being the spacetime co-ordinates. \\ 

The normal cosmological background of the $C$ - field is below this threshold.  Had the universe been homogeneous, there 
would have been no creation of matter.  However, the real universe, although smooth and homogeneous on a large enough 
scale (so that it can be described by the Robertson - Walker line element), has pockets of strong gravitational field, 
such as in the neighbourhood of collapsed massive objects, often dignified by the name `black holes'. \\

We shall use the name `near black hole' (NBH) to denote a collapsed massive object whose size is very slightly in excess 
of that of its event horizon, {\it if it were a Schwarzschild black hole}.  Thus a spherical object of mass M would have a radius 

\begin{equation}
R=\frac{2GM}{c^2}+\epsilon, ~~~~~~~\epsilon \ll R.
\end{equation}

\noindent In the neighbourhood of such an object, at a distance $r$ from its centre,
\begin{equation}
C_iC^i=\frac{{m_c}^2}{1-\frac{2GM}{c^2r}},~~~~r>R 
\end{equation}

\noindent where ${m_c}( < m_p)$ is the background level of the $C$ - field energy density.  It is thus possible that at $r$ 
sufficiently close to $R$, the value of $C_iC^i$ crosses the creation threshold.  This is when creation of matter would take 
place.  And, because the creation of matter is accompanied by the creation of the $C$ - field, the latter generates negative 
stresses close to $r = R$, which blow the created matter outwards. \\

In the above example the creation is isotropic and the resulting disturbances will not generate gravitational waves.  
This situation is, however, highly idealized.  The real massive object will not be spherically symmetric, nor would the 
creation and expulsion of new matter from it be isotropic about its centre. \\ 

Even the next stage of asymmetry is sufficient to generate gravitational waves. namely that of a spinning collapsed massive 
object which is idealized as the Kerr black hole.  The line element of the external spacetime for such a black hole is given by 
\begin{equation}
ds^2=\frac{\Delta}{\rho^2}(dt-h{\rm sin}^2\theta d\phi)^2-\frac{{\rm sin}^2\theta}{\rho^2}[(r^2+h^2)d\phi-hdt]^2-\frac{\rho^2}{\Delta}dr^2-\rho^2d\theta^2,
\end{equation}
\noindent where
\begin{equation}
\Delta \equiv r^2-2mr+h^2,~~~\rho^2 \equiv r^2+h^2{\rm cos}^2\theta,
\end {equation}
\noindent and $m=M G/c^2$, $h=J/M c$ where, $M$ and $J$ are respectively the mass and angular 
momentum of the black hole. The Kerr black hole has an outer horizon at 

\begin{equation}
r_+=m+\sqrt{m^2-h^2},
\end{equation}

\noindent while the surface of revolution given by,
\begin{equation}
r_s(\theta)=m+\sqrt{m^2-h^2{\rm cos}^2\theta},
\end{equation}
\noindent is called the static limit.  Between $r_+$ and $r_s$ is the region known as the ergosphere, wherein matter is made to 
rotate in the same direction as the spinning black hole. \\

As in the case of the Schwarzschild Black Hole the spinning collapsed massive object here will simulate a near black hole 
with spin and its exterior solution will be approximately given by the above equation(7).  Here too we expect the ${C_i}{C^i}$ 
to be raised above the threshold close to the horizon. For, it is given by

\begin{equation}
C_iC^i=\frac{[(r^2+h^2)(r^2+h^2{\rm cos}^2\theta )+2mrh^2{\rm sin}^2\theta]}{(r^2+h^2{\rm cos}^2\theta ) (r^2-2mr+h^2)}.
\end{equation}
\noindent It can be seen that the above expression is maximum at poles $(\theta=0,\pi)$ and minimum at the equator $(\theta=\pi/2)$.  
Thus the creation threshhold will be attained more easily at the poles than at the equator, leading to preferential creation of 
matter there. However, because of the ergosphere property of dragging any matter along with the spinning mass, only matter 
created near the polar regions $(\theta = 0, \pi)$ would find it way out as it is ejected by the $C$ - field. {\it In other 
words we expect created matter to find its way out along the polar directions in the form of oppositely directed jets.}  
This is the canonical source of gravitational waves in the QSSC. \\

A word of caution is needed in the above argument.  We have assumed that in the neighbourhood of a typical NBH, the strength 
of the cosmological $C$ - field will be small.  Thus we have assumed that the Schwarzschild and Kerr solutions are not significantly 
modified by the $C$ - field.  This assumption can be checked only by obtaining an exact solution of a NBH with the $C$ - field.  
We have not carried out this (rather difficult, possibly impossible) exercise; but have relied on approximations based on 
series expansions.  In any case for the purpose of this paper, we have given a rationale for expecting the simplest cosmological 
sources of gravitational waves as twin jet systems spewing out newly created matter linearly in opposite directions. 

\section{Gravitational radiation from a mini creation event}
\indent
In the QSSC the created matter near a Kerr-like black hole moves rapidly along the polar directions in the 
form of oppositely directed jets. Such an object is endowed with a changing quadrupole moment causing the system to emit 
gravitational waves. In the following we estimate its amplitude. \\

We set up a spherical polar co-ordinate system in which the jet is expanding linearly with a speed $u$ in $\ep$-$\psi$ 
direction and consider the $z$-direction to be the line of sight. The gravitational wave amplitude under quadrupole approximation 
(a good approximation for a very long distance source) can be calculated from the reduced mass quadrupole moment of the 
source. The reduced mass quadrupole moment of a source is given by

\begin{equation}Q_{ij}=\int_{V} \rho dV \left(r_{i}r_{j}-\frac{1}{3}\delta_{ij}r^2\right), \end{equation}
where $\rho$ is the mass per unit volume. We assume that away from the near black hole the geometry is almost Euclidean. \\
\indent
The components of gravitational wave amplitude at detector time $t$ are given by
\be\bar{h}_{ij}=\frac{2G}{c^4 R}\left[\ddot{Q}_{ij}\left(t-\frac{R}{c}\right)\right], \ee 
where $R$ is the radial distance of the object from the detector. At the source we have the time $t_0=t-R/c$.
The transverse traceless components can be extracted from $\bar{h}_{ij}$ through the projection operator $P_{a}^{b}=\delta_{a}^{b}-n^b n_{a}$ as

\be \bar{h}_{ij}^{TT}=P_i^k P_j^l \bar{h}_{kl}-\frac{1}{2} P_{ij}\left(P^{kl}\bar{h}_{kl}\right). \ee

For gravitational waves propagating along z-direction, $n_a=(0,0,1),$ the `$+$' and the `$\times$' polarisation components of the wave are 
\be h_+=\frac{\bar{h}_{11}-\bar{h}_{22}}{2} \mbox{  and   } h_\times=\bar{h}_{12}. \ee
We will assume the radial velocity of the jet to be $u$ so that $|ut_0|<< R$.\\
For a time $t_{0}$ typical of the source, the three spatial co-ordinates are $r_{1}=ut_{0}\, \rm sin\ep\, \rm cos\psi$ , $r_{2}=ut_{0}\, \rm sin\ep\, \rm sin\psi$ 
and  $r_{3}=ut_{0}\, \rm cos\ep.$ The non-vanishing components of the symmetric mass quadrupole moment ($Q_{ij}$) tensor are : 

\begin{eqnarray} 
Q_{11}&=&\frac{2 Q_{0}}{3} \left(\,\rm sin^2\ep \, \rm cos^2\psi - \frac{1}{3}\right),\nonumber\\
Q_{12}&=&\frac{Q_{0}}{3}\, \rm sin^2\ep\, \rm sin2\psi, \nonumber\\
Q_{13}&=&\frac{Q_{0}}{3} \rm sin2\ep\, \rm cos\psi, \nonumber \\
Q_{22}&=&\frac{2 Q_{0}}{3} \left(\,\rm sin^2\ep\, \rm sin^2\psi - \frac{1}{3}\right), \nonumber\\
Q_{23}&=&\frac{Q_{0}}{3} \rm sin2\ep\, \rm sin\psi, \nonumber\\
Q_{33}&=&\frac{2 Q_{0}}{3} \left(\,\rm cos^2\ep - \frac{1}{3}\right),
\end{eqnarray}
\noindent where $Q_{0}=\dot{M} u^{2} t_{0}^3$. The mass creation rate is $\dot{M}=A\rho u$, where $A$ is the area of the jet.
\vskip 0.5cm
The two polarisation components can be calculated as
\begin{eqnarray}
h_+(t_0)&=&h_{char} \,\rm sin^2\ep \, \rm cos2\psi, \nonumber\\
h_\times(t_0)&=&h_{char} \, \rm sin^2\ep \, \rm sin2\psi,
\end{eqnarray}
where
\begin{eqnarray}
h_{char}&=&\frac{4 G \rho A u^3 t_0}{c^4 R} \nonumber \\
        &\sim &2.7\times 10^{-17} \left (\frac{\dot{M}}{200 ~M_{\odot}/\rm {sec}}\right ) \left (\frac{{\it{u}}}{0.8 {\it{c}}}\right ) \left (\frac{t_0}{1000 ~\rm {sec}}\right ) \left (\frac{{\it{R}}}{3 ~\rm {Gpc}}\right )^{-1}. \nonumber \\
\end{eqnarray}
The mini creation event sweeps over the band of the detector from high to low frequency. For LISA the low end of the band is taken
to be $10^{-4} Hz$ which corresponds to about $10^4 \rm {sec} \sim$ few hours. In this period the LISA hardly changes orientation.
However, assuming random distribution and orientation of the MCE we perform appropriate averages over the directions and orientation
of the MCE. \\       

The angle averaging of a quantity $\it$V is performed according to 
\begin{eqnarray}
<V>_\psi &=&\left[\frac{1}{2\pi}\int_{0}^{2\pi} V^2 d\psi \right]^\frac{1}{2}, \nonumber\\
<V>_\epsilon &=&\left[\frac{1}{2}\int_{0}^{\pi} V^2 \rm sin\epsilon d\epsilon \right]^\frac{1}{2},\nonumber\\
<V>_{\epsilon,\psi}&=&\left[\frac{1}{4\pi}\int_{\psi=0}^{2\pi}\int_{\epsilon=0}^{\pi} V^2 \rm sin\ep d\ep d\psi \right]^\frac{1}{2}. \nonumber
\end{eqnarray}
Thus,
\begin{eqnarray}
<h_+(t_0)>_\psi&=&<h_\times(t_0)>_\psi ~~= ~~\frac{1}{\sqrt{2}} \, h_{char} \, \rm sin^2\epsilon , \nonumber\\
<h_+(t_0)>_{\epsilon,\psi}&=&<h_\times(t_0)>_{\epsilon,\psi} ~~= ~~\frac{2}{\sqrt{15}} \, h_{char}. \nonumber\\
\end{eqnarray}	       
\section{Fourier Transform of the gravitational wave \\ amplitude}

The two polarisations of gravitational wave amplitude that we have found in the earlier section have to be expressed 
in a limited frequency space within which the detector is supposed to be most sensitive. Also here we redefine our zero of 
time as the instant when the gravitational radiation first hits the observer. \\

The Fourier transform of $h_+(t_0)$ and $h_\times(t_0)$ are given by 

\begin{eqnarray} 
\tilde{h}_+(f)&=&\int_{-\infty}^{\infty}h_+(t_0) \, exp \, (-2 \, \pi \, \it i \, f \, t_0) \, dt_0 \nonumber\\
              &=&\frac{\dot{M} G u^2}{\pi^2 c^4 R}\frac{1}{f^2}\,\rm sin^2\ep\,\rm cos2\psi. \\
\tilde{h}_\times(f)&=&\int_{-\infty}^{\infty}h_\times(t_0)\,exp(-2\pi\,\it i\,f\,t_0)\,dt_0 \nonumber\\
                   &=&\frac{\dot{M} G u^2}{\pi^2 c^4 R}\frac{1}{f^2}\,\rm sin^2\ep\,\rm sin2\psi.
\end{eqnarray}
\begin{figure}
\epsfbox{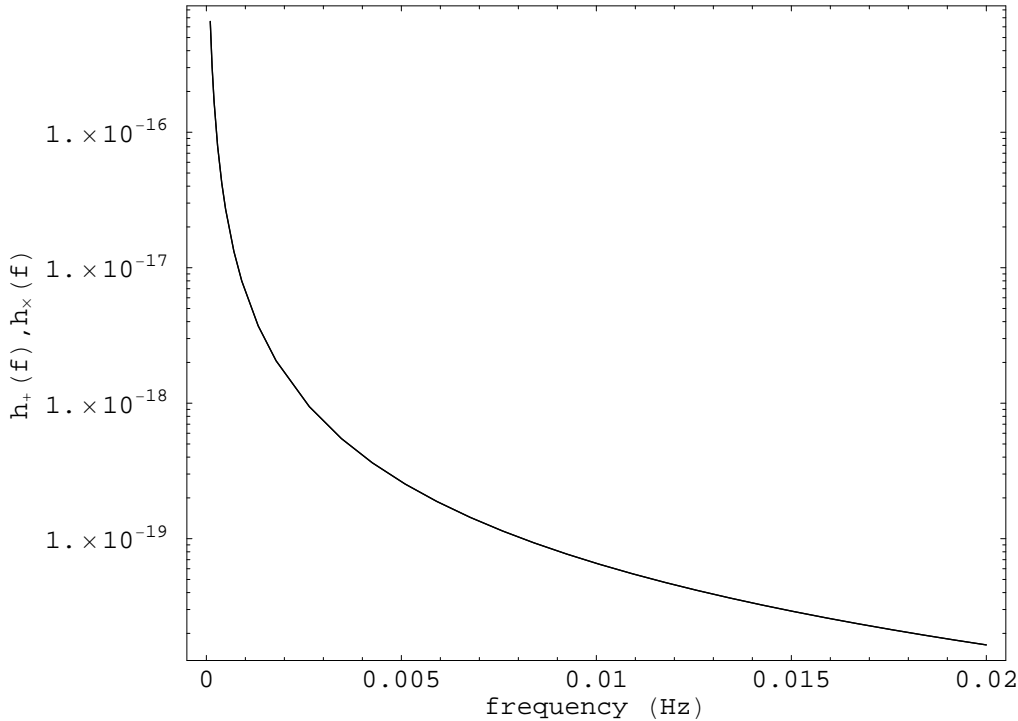}
\vspace*{5mm}
\caption{The Fourier Transform of the two polarisations of the wave, $\rm h_+(\rm f)$, $\rm h_\times (\rm f)$ in units of 
$Hz^{-1}$ are plotted here as a function of frequency for a MCE jet with $\dot{M}=200 M_{\odot}/\rm {sec}$, $R=c/{2 H_0}\quad (\textrm{for~~} h_0=0.65)$, $u=0.8 c$. 
Here $\ep \sim 15^0$ and $\psi$ is averaged over uniformly. The two components thus averaged are of the same amplitude and hence they overlap.}
\end{figure} \\
Plots of $\tilde{h}_+(f)$ and $\tilde{h}_\times(f)$ are shown in Figure 1.
Since the frequency $\it f $ appears in the denominators of the above expressions, the amplitude of the jet falls in inverse 
square fashion as it expands. This causes most of the wave energy to be concentrated in the low frequency bin. It is thus 
worthwhile to analyse the detectibility of such a wave through low frequency detector such as LISA. We will also compare the results with the magnitude detectable by the high frequency detector like LIGO. \\
\section{Detectibility of MCE by LISA capability}
\subsection{Time Delay Interferometry of LISA}
LISA, the Laser Interferometric Space Antenna, is a proposed mission that will use coherent laser beams exchanged between 
three identical spacecraft forming a giant equilateral triangle with each side $5\times10^6$ km to observe and detect low 
frequency gravitational waves from cosmic sources. \\

In LISA six data streams arise from the exchange of laser beams between the three spacecraft. The sensitivity of LISA crucially 
depends on the cancellation of the laser frequency noise. It is because of the impossibility to achieve equal distances between 
spacecraft, the laser frequency cannot be exactly cancelled to enhance its sensitivity. Several schemes came up to combine 
the recorded data with suitable time delays corresponding to the three arm lengths of the giant triangular interferometer. \\

The idea of time delayed data combination scheme was proposed by the Jet Propulsion Laboratory team (1999a, 1999b, 2000). 
Dhurandhar et. al.(2002) adopted an algebraic approach to this problem of introducing time delays to cancel the laser frequency 
noise based on the modules of polynomial rings.\\

The data combinations that cancel the laser frequency noise consist of six suitably delayed data streams, the delays being 
integer multiples of the light travel times between spacecraft, which can be conveniently expressed in terms of polynomials 
in the three delay operators $E_1,E_2,E_3$ corresponding to the light travel time along the three arms. The laser noise 
cancellation condition puts three constraints on the six polynomials of the delay operators corresponding to the six data 
streams. The problem therefore consists of finding six-tuples of polynomials that satisfy the laser noise cancellation 
constraints. These polynomial tuples form a module, called  the {\it module of syzygies}. \\ 

Given any elementary data streams $U^i, V^i$, a general data combination is a linear combination of these elementary data streams

\be X(t)=\sum_{i=1}^{3} p_i V^i (t)+q_i U^i (t) \ee
where $p_i$ and $q_i$ are polynomials in the time-delay operators $E_i,~i=1,2,3$. Thus any data combination can be expressed as 
a six-tuple polynomial `vector' ($p_i$,$q_i$). For cancellation of laser frequency noise only the polynomial vectors satisfying 
this constraint are allowed and they form the module of syzygies mentioned above. While the laser frequency noise and optical 
bench motion noise can be canceled by taking appropriate combinations of the beams in the module of syzygies, the acceleration 
noise of the proof masses and the shot noise cannot be canceled out in the scheme. These then form the bulk of the noise spectrum. The noise power spectral density is also expressible in terms of 
the noise cancelling polynomials of the time delay operators.  For different combinations, the expression for the noise 
spectrum will also be different. \\

In our analysis we use the Michelson combination to calculate the response and the noise power spectral density. As shown 
in Nayak et. al. (2003), the Michelson combination on the average has almost as good sensitivity as the optimized 
combinations. Since here we are interested in order of magnitude estimates, the Michelson combination is good enough for 
our purpose. Moreover, it is easier to calculate relevant quantities for the Michelson combination than for other combinations.   

\subsection{Estimation of Signal-to-Noise Ratio of jets in LISA}
We choose a co-ordinate system in which the LISA configuration is at rest and let x-axis of the co-ordinate system be 
perpendicular to one of the LISA arm unit vectors ($\hat n_1, \hat n_2, \hat n_3$). The z-axis is considered perpendicular 
to the plane of the LISA triangle. The unit vector $\hat w$ connecting the origin and the source is parametrized by the source angular location ($\theta, \phi$), so that 

\be \hat w =\left(\begin{array}{ccc} 
		\rm sin\theta~\rm cos\phi \\
		\rm sin\theta~\rm sin\phi\\
		\rm cos\theta \end{array} \right) 
\ee

and the transverse plane is spanned by the unit transverse vector $\hat \theta$ and $\hat \phi$,~defined by
\be \hat \theta =\frac{\partial \hat w}{\partial \theta}~,\qquad \hat \phi = \frac{1}{\rm sin\theta} \frac{\partial \hat w}{\partial \phi}~. \ee

For Michelson combination we have the expression of the polynomial 

\begin{eqnarray} 
p_i&=&\{1 - E_2~E_2, 0, E_2~(E_3~E_3 - 1)\} \nonumber\\
q_i&=&\{1 - E_3~E_3, E_3~(E_2~E_2 - 1), 0\} \nonumber\\
r_i&=&\{E_2~E_2 + E_3~E_3 - E_2~E_2~E_3~E_3 - 1, 0, 0\}
\end{eqnarray}
$E_i=e^{{\it i}\Omega L_i}$ and $\Omega=2\pi f$ is the angular frequency, where the LISA arm vectors are $\vec r_i=L_i~\hat n_i ~(i=1, 2, 3)$. However, 
for purpose of calculation of the noise and response we can consider $L_i$ to be equal to \rm {L} say.\\ 

The noise power spectral density as given by Bender et. al. (2000) for this combination is 
\be
S_{\it {M}}=16~S_{\it {shot}}~{\rm {sin}}^2 2\pi {\it{f}} {\it{L}}~+32~S_{\it {proof}}(2~{\rm {sin}}^2 2\pi {\it{f}} {\it{L}}~-{\rm {sin}}^4 2\pi {\it{f}} {\it{L}}),
\ee
where, \\
\begin{eqnarray}
S_{\it shot}&=&5.3\times 10^{-38}f^2~ Hz^{-1}~~,\nonumber  \\
S_{\it proof}&=&2.5\times 10^{-48}f^{-2}~ Hz^{-1}~~.\nonumber
\end{eqnarray}
Plot of the noise power density $S_{\it M}$ for Michelson combination is shown in Figure 2. Several sets of generators have 
been listed in Dhurandhar et. al. (2002). The response of LISA to a source is expressible for a given data combination $X$ 
in terms of its elementary data stream  $U^i, V^i$ as the following :
\be {\cal R}_X(\Omega ;  \theta, \phi, \ep,\psi) =\tilde h_+(\Omega ;  \ep, \psi)F_+(\Omega ;  \theta, \phi)+\tilde h_\times(\Omega ;  \ep, \psi)F_\times(\Omega ;  \theta, \phi) \ee
where $F_{+,\times}(\Omega)$ are transfer functions which correspond to the combination $X$. These are functions of the source 
angular location and the frequency. For different noise cancelling combinations $U_i, V_i, F_{+,\times}(\Omega)$ will have different 
expressions. Below are quantities for  $U_1,V_1$ (the others are obtained by cyclic permutations) :

\begin{eqnarray}
F_{U_1 ; +, \times}&=&\frac{e^{\it i \Omega(\hat w. \vec r_3 +L_2)}}{2(1+\hat w. \hat n_2)}\times(1-e^{-\it i\Omega L_2 (1+\hat w. \hat n_2)}) \xi_{2;+,\times} ~, \nonumber \\
F_{V_1 ; +, \times}&=&-\frac{e^{\it i \Omega(\hat w. \vec r_2 +L_3)}}{2(1-\hat w. \hat n_3)}\times(1-e^{-\it i\Omega L_3 (1-\hat w. \hat n_3)}) \xi_{3;+,\times} ~,
\end{eqnarray}
where,
\begin{eqnarray}
\xi_{i;+}&=&(\hat \theta .\hat n_i)^2-(\hat \phi .\hat n_i)^2 \;,\nonumber\\
\xi_{i;\times}&=&2(\hat \theta .\hat n_i)(\hat \phi .\hat n_i).
\end{eqnarray}
In the above $L_i$ are LISA arm lengths (i=1,2,3), $\hat w$ is the unit vector along the line of sight, $\hat \theta$ and $\hat \phi$ 
are the unit vectors transverse to the line of sight.  \\ 
\begin{figure}
\epsfbox{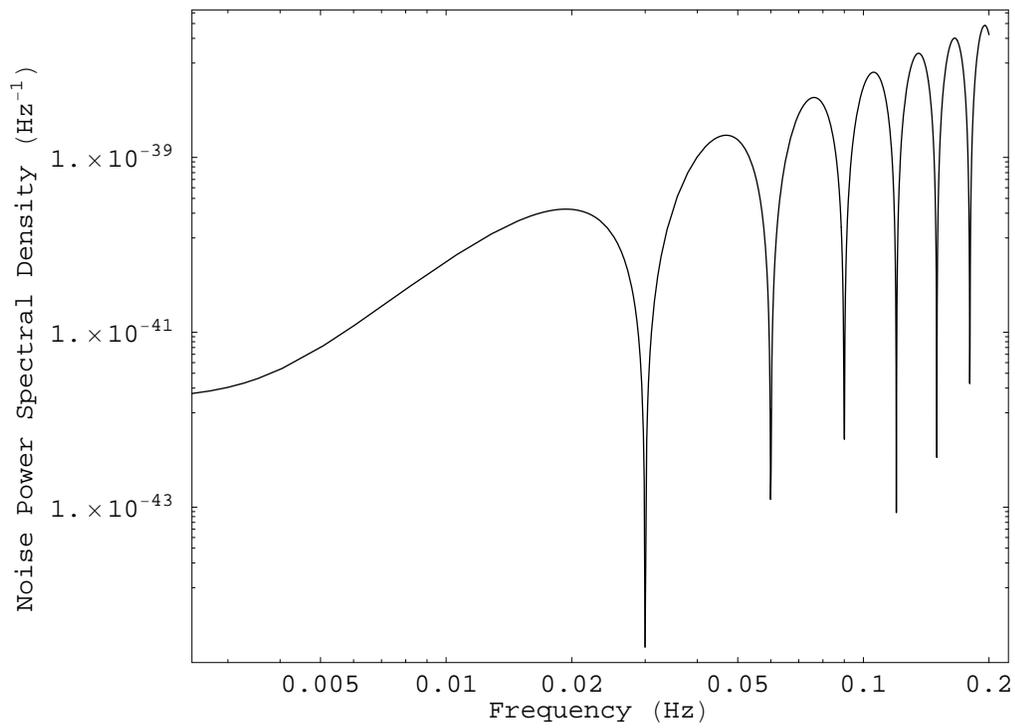}
\vspace*{5mm}
\caption{Noise power spectral density $S_{\it M}$ for Michelson combination, as a function of frequency.}
\end{figure} \\
The response of LISA to Michelson combination is given by
\be
{\cal R}_{\it M}=\sum_{\it i=1}^3 [p_i (F_{V\it i ;+}h_+ +F_{V\it i ; \times}h_{\times})+q_i (F_{U\it i ;+}h_+ +F_{U\it i ; \times}h_{\times})]~, \ee
where the polynomial functions $p_{i}$ and $q_{i}$ are for the Michelson combination given in equation (22).
The signal-to-noise ratio corresponding to a particular frequency, say $f$, is given by
\be SNR_f = \frac{|{\cal R}_{\it{M}}|}{\sqrt{S_{\it{M}}}}. \ee
The integrated signal-to-noise ratio is then given by
\be SNR= \left [ 2 \int_{0}^{\infty} \frac{|{\cal R}_{M}|^2}{S_{M}} ~ df \right ]^ \frac{1}{2}. \ee

The response ${\cal R}_M$ is now a function of $\ep, \psi, \theta, \phi$ and the frequency $f$. Since the detector is omnidirectional, 
it will pick up all such sources of varying angles and frequency and give the output. So, one needs to average the response for 
those angles before plugging it in the formula for the signal-to-noise ratio calculation. The expression for the integrated 
SNR takes care of the frequency averaging. \\

We apply this model to a superluminal jet beaming towards us nearly at an angle $15^o$ from a distance of about half the 
Hubble distance and with a speed of nearly 80\% of the speed of light in vacuum, i.e. in this case,
$ u=0.8 c$, \, $\ep=15^o$, \, $R=c/{2 H_0}\quad (\textrm{for~~} h_0=0.65)$. \\
\begin{figure}
\epsfbox{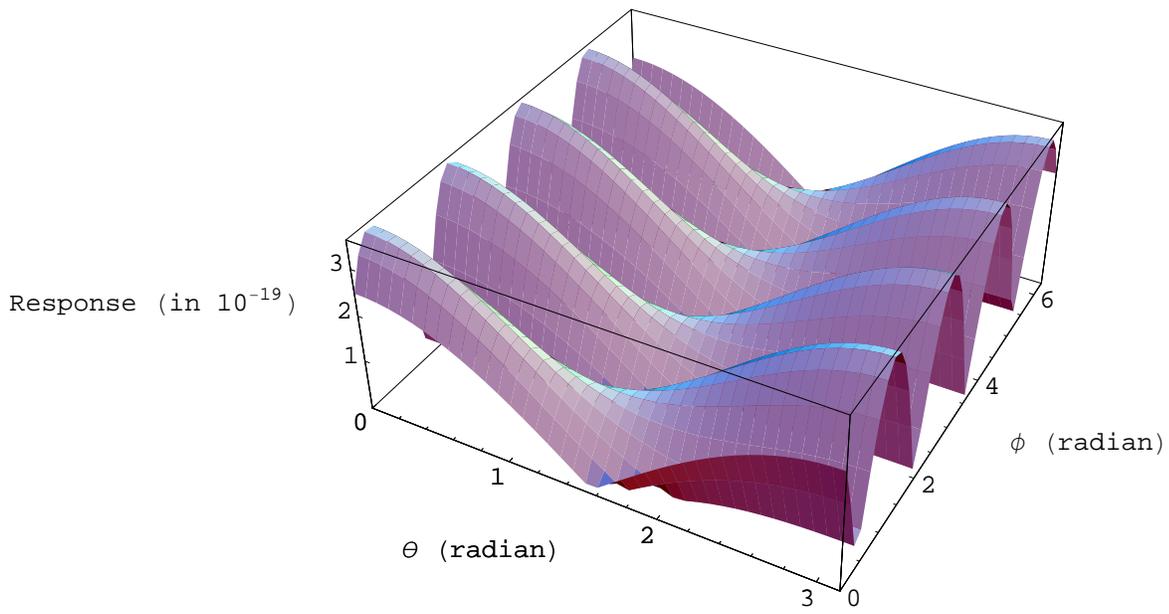}
\vspace*{5mm}
\caption{Response $\cal R_{\it M}$ of LISA to the jet source at $f=1 ~m Hz$ as function of angular location ($\theta$, $\phi$) of the source for the Michelson combination.}
\end{figure} \\
Plot of $\cal R_{\it M}$ of LISA for the Michelson combination is shown in Figure 3. We have implemented the response function given by 
Dhurandhar et al. (2002) and their proper angle averaging and calculated the signal-to-noise-ratio as a function of frequency and the integrated signal-to-noise ratio over the frequencies in the 
LISA window. Because the sensitivity of LISA window starts `bumping' in the higher frequency side, and also because the 
jet source has comparatively lower amplitude in this region and thus contributes less to the integrated signal-to-noise 
ratio, we consier a frequency range $10^{-4}~-~2\times10^{-2}$ Hz for LISA sensitivity to calculate the relevant quantities. \\

%\begin{tabular}{|c|c|}
%\hline
%                 &                       \\
%{\bf $\frac{\dot{M}}{M_{\odot}}$}	& {\bf Integrated SNR}	\\
%                &                       \\
%\hline
%                &			\\
%$ 10^2$		& $5.1958$		\\
%		& 			\\
%\hline
%		&			\\
%$10^3$		& $51.958$		\\
%		&			\\
%\hline
%                &                       \\
%$10^4$		& $519.584$		\\
%		& 			\\
%\hline
%                &                       \\
%$10^5$		& $5195.84$		\\
%		&			\\
%\hline
%                &                       \\
%$10^6$		& $51958.4$		\\
%		&			\\
%\hline
%                &                       \\
%$10^7$		& $519584$		\\
%		&			\\
%\hline
%\end{tabular}

From the results, it is apparent that SNR scales linearly with the mass creation rate as the following:

\be SNR=5.19 \times \left( \frac{\dot M}{100~M_\odot/\rm sec} \right ) .  \ee 

This is also obvious from the expression for the gravitational wave amplitude. From this expression, the minimal mass creation 
rate for which a source is just observable, can be calculated. Here we consider this `bare visibility' value for SNR to be 
$~10$. In this case the mass creation rate turns out to be about  $200 M_{\odot}/\rm{sec}$. \\ 

The net gravitational wave amplitude from an MCE, in the linear approximation, can be thought of as arising form the 
superposition of gravitational waves from individual fluid elements of the MCE. If $f$ is the frequency of interest, 
coherent superposition of amplitudes occurs only from within a region $2\cal L$ of size less than $~0.5\lambda=0.5c/f$. 
For the LISA window $(f_1=10^{-4} \: - \: f_2=2\times 10^{-2})Hz$, we have

\begin{eqnarray}
{2\cal L}_1 <\frac{c}{2\times 10^{-4} Hz} \qquad &\Rightarrow& \quad {\cal L}_1< (2500  ~\rm seconds)\times c \nonumber\\
{2\cal L}_2 <\frac{c}{2\times 2~10^{-2} Hz} \qquad &\Rightarrow& \quad {\cal L}_2< (12.5  ~\rm seconds)\times c \nonumber
\end{eqnarray}
{\it {i.e.}}, towards higher frequency side the jet length limit is smaller and towards lower, it is larger. This means, 
under the linear approximation, the gravitational wave frequency in LISA window will carry information about the MCE during 
which jet acquires the above length limit. This length limit can be converted to time limit from the knowledge of the velocity 
of expansion of the jets.\\ 
The corresponding time limit is :
$$\left(\frac{{\cal L}_1}{u}-\frac{{\cal L}_2}{u} \right)\,=\,\left (\frac{2500c}{0.8c}-\frac{12.5c}{0.8c}\right ) ~\rm {seconds}\,=\,3109.375~\rm seconds \approx ~52~\rm {minutes},$$
{\it {i.e.}}, LISA observation of MCE can look into the history from 15.625 seconds after the birth of an MCE for a duration of about 52 minutes. 
But LISA will be able to `see' the jets for which the mass creation rate is at least ${\dot M}=200 M_{\odot}/\rm{sec}$. 
This mass creation rate then gets converted to a single jet mass window of $(1.56\times 10^{3},~~3.13\times 10^{5}) \, M_{\odot}/\rm {sec}$.
\section{Observation of MCE by LIGO type detector}
For the LIGO type detectors Wiener optimal filters $q(t)$ are used. These filters are defined by their Fourier transform(Thorne 1987, Schutz 1991) as
\begin{equation}
\bar{q}(f)=k\frac{\bar{h}(f)}{S_n(f)},
\end{equation}
where $S_n(f)$ is the spectral density of the noise in the detector and ${\it{k}}$ is some arbitrary constant. For a laser 
interferometric detector of advance LIGO type $S_n(f)$ is given in the following form
\begin{equation}
S_n(f) = \left\{ \begin{array}{ll}
\infty,                                                   &    f<10H_z ;\\ 
S_0[{(\frac{f_0}{f})}^4 + 2\{1+(\frac{f^2}{{f_0}^2})\}],    &    f>10 H_z,
\end{array}
\right.
\end{equation} 
where $S_0 = 3\times 10^{-48} {H_z}^{-1}$ and $f_0 = 70 H_z$ (Cutler and Flanagan, 1994 \footnote{Since we are only interested in the order of magnitude estimated, this noise curve suffices}). The amount of detector noise 
determines the strength of the weakest detectable signal by the detector. For a perfect frequency matching of the filters 
with the signal, the cross-correlation between the detector outputs and the filter leads to a signal to noise ratio
\begin{equation}
\frac{S}{N}=\left[2 \int^{\infty}_{0}\frac{|\bar{h}(f)|^2}{S_n(f)} df\right]^{1/2}.
\end{equation}

We assume that the present detector is uniformly sensitive in the frequency band $10H_z$ to $1000H_z$ and is blind everywhere 
outside it. Then equation(16) reduces to the form
\begin{equation}
\frac{S}{N}=\frac{\sqrt{2} G \dot{M} u^2 f_0}{3 c^4 {\pi}^2 R {S_0}^{1/2}}\left[\int^{1000 H_z}_{10 H_z}\frac{df}{(2 f^6+2 f_0^2 f^4+f_0^6)}\right]^{1/2},
\end{equation}
Integrating the above integral numerically, we get
\begin{equation}
\frac{S}{N}\sim \frac{G \dot{M} u^2}{c^4 {\pi}^2 R}(3.76\times 10^{20}) {H_z}^{-1}.
\end{equation}
On choosing the following values : \\ 
\begin{eqnarray}
r&=&\frac{c}{2 H_0}\equiv {\rm half ~the ~Hubble ~distance} ~({\rm for} ~h_0=0.65), \nonumber \\ 
u&=&\frac{8c}{10}\equiv ~{\rm eight ~tenth  ~of  ~the  ~velocity  ~of  ~light}, \nonumber \\ 
\dot{M}&=&\frac{\kappa}{\tau} M_{\odot}, ~\kappa ~{\rm is ~a ~constant ~and }~\tau ~{\rm is ~the ~proper ~time ~measured ~in ~seconds}
\end{eqnarray}
we obtain
\begin{eqnarray}
\frac{S}{N}&\sim &5.3\times 10^{-4}\left(\frac{\dot{M}}{M_{\odot}}\right) \nonumber \\ 
&=&5.3\times 10^{-4}\left(\frac{\kappa}{\tau}\right).
\end{eqnarray}

We next relate such values to actual situations, for which we need to go back to the astrophysics of a mini-creation event.
We return to Section 2 to discuss further details of the typical mini-creation event (MCE).  The $C$-field obeys the wave 
equation with sources in the world points where particle creation takes place (Hoyle, et al 1995).  So when matter is created 
at a point, the negative energy C-field tends to escape outwards more efficiently than the matter created, which has nonzero 
restmass. Thus initially the mass created adds to the existing massive object.  It is this tendency that leads to build up 
of massive concentrations of matter such as found in the nuclei of galaxies.  However, as the central mass grows and its 
gravitaional field increases in strength, the free escape of the $C$-field quanta is inhibited and this leads to a concentration 
of the field in the object.  Since the field has negative stress, the interior of the collapsed object tends to become unstable.  
With sufficient accumulation of $C$-field strength, it may break up and cause some pieces to be thrown out with great speed.
It was this scenario that was envisaged in Section 2 and as stated there, if the collapsed object is a near-Kerr black hole, 
it will eject material along the axis.  Although the material is coming out of a region of high gravitational redshift, its 
ejection speed can be  even more dominant  and allow it to come out with high speed.  The situation is somewhat like the 
classical white hole (see for example Narlikar, et al 1974), except that in this case the $C$-field is the driving agent which 
prevents the outgoing material from being swamped out by the relative inward motion of the surrounding material, thus 
overcoming an objection to white holes envisaged by Eardley (1974). In such a case, the early expansion is very rapid, with
the external observer receiving radiation that is highly blueshifted. The blue shift does not last long, however, and the 
expansion slows down subsequently.
\section{Conclusions}
The gravitational waves could be generated in a chain of endless mini bangs if there is a small anisotropy present in the 
process. An anisotropic mini creation event is the biggest source of the gravitational waves. The calculations we performed 
here show that a laser interferometric detector of the LISA type can be used to detect through a window of low frequency range 
$10^{-1}Hz$ to $2\times10^{-2}Hz$ for a duration of about 52 seconds. In this duration LISA will be able see 'jets' for which 
the mass creation rate is at least $200M_\odot/\rm{sec}$. Whereas a laser interferometric detector of the advance LIGO type  
can be used to detect the mini creation events, which opens its window of high frequency range $10Hz$ to $10^3Hz$ for a very 
short duration of the order of $10^{-2}$ seconds and in this duration it observes 'jets' for which the mass creation rate 
is $2\times10^4M_\odot/\rm{sec}$. It appears from this elementary analysis that the LISA detector is well suited to detect 
MCEs through their gravitational waves while the LIGO may have a less sensitive, marginal role to play. \\ 
\vspace{5mm}
\begin{flushleft}
{\large{\bf{References}}} \\ 
\vspace{5mm} 

\small{Armstrong J.W., Eastabrook F. B., Tinto M., 1994, ApJ, 527, 814}. \\ 

\small{Bender P.{\it et al.}, 2000, `LISA: A Cornerstone Mission for the Observation of Gravitational Waves', \\~~~~~~~~~~System and Technology Study Report ESA-SCI (2000), 11}. \\

\small{Bondi H., Gold T., 1948, MNRAS 108, 252}. \\

\small{Cutler C., Flanagan E. E., 1994, Phys. Rev. D 49, 2658}. \\

\small{Das Gupta P., Narlikar J. V., 1993, MNRAS 264, 489}. \\ 

\small{Dhurandhar S. V., Nayak K. Rajesh, Vinet J-Y., 2002, Phys. Rev. D 65, 102002}. \\

\small{Eardley, D.M. 1974, preprint}. \\

\small{Eastabrook F. B., Tinto M., Armstrong J.W., 2000, Phys. Rev. D 62, 042002}. \\

\small{Hoyle F., 1948, MNRAS 108, 372}. \\

\small{Hoyle F., Burbidge G., Narlikar J. V., 1993, ApJ 410, 437}. \\

\small{Hoyle F., Burbidge G., Narlikar J. V., 1994a, MNRAS 267, 1007}. \\

\small{Hoyle F., Burbidge G., Narlikar J. V., 1994b, A\&A 289, 729}. \\

\small{Hoyle F., Burbidge G., Narlikar J. V., 1995a, Proc. Roy. Soc. A 448, 191}. \\ 

\small{Hoyle F., Burbidge G., Narlikar J. V., 2000, `A Different Approach to Cosmology', Cambridge \\~~~~~~~~~~Univ. Press, Cambridge}. \\

\small{Narlikar J.V., Apparao, K.M.V., Dadhich, N.K., 1974, Nature, 251, 590}. \\

\small{Narlikar J.V., Padmanabhan T, 2001, Annu. Rev. A\&A, 39, 211}. \\

\small{Narlikar J.V., Vishwakarma R.G., Burbidge G., 2002, PASP, 114,1092}. \\

\small{Nayak K. Rajesh, Dhurandhar S.V., Pai A., Vinet J-Y., 2003, Phys. Rev. D 68, 122001}. \\

\small{Sachs R., Narlikar J. V., Hoyle F., 1996, A\&A 313, 703}. \\

\small{Schutz B. F., 1991, in Blair D. G., ed., The Detection of Gravitational Waves. Cambridge Univ. \\  ~~~~~~~~~~Press, Cambridge, p. 406}. \\ 

\small{Thorne K. S., 1987, in Hawking S., Israel W., eds., 300 Years of Gravitation. Cambridge Univ. \\  ~~~~~~~~~~Press, Cambridge, p. 330}. \\

\small{Tinto M., Armstrong J.W., 1999, Phys. Rev. D 59, 102003}. \\

\end{flushleft}
\end{document}